\documentclass[reprint,
superscriptaddress,
 amsmath,amssymb,
aps,
prl,
]{revtex4-2}

\usepackage{graphicx}
\usepackage{dcolumn}
\usepackage{bm}
\usepackage{MnSymbol}
\usepackage{xcolor}

\usepackage[normalem]{ulem}

\begin{document}


\title{The random force in molecular dynamics with electronic friction}

\author{Nils Hertl}
\affiliation{%
 Max-Planck-Institut f\"ur biophysikalische Chemie, Am Fa\ss berg 11, 37077 G\"ottingen, Germany}
\affiliation{Institut f\"ur physikalische Chemie, Georg-August-Universit\"at G\"ottingen, Tammanstra\ss e 6, 37077 G\"ottingen, Germany}

\author{Raidel Martin-Barrios}
\affiliation{Universit\'e de  Bordeaux, 351 Cours de la Lib\'eration, 33405 Talence, France}
\affiliation{CNRS, 351 Cours de la Lib\'eration, 33405 Talence, France}
\affiliation{Universidad de La Habana, San L\'{a}zaro y L, CP 10400 La Habana, Cuba}

\author{Oihana Galparsoro}%
\affiliation{%
 Max-Planck-Institut f\"ur biophysikalische Chemie, Am Fa\ss berg 11, 37077 G\"ottingen, Germany}
\affiliation{Institut f\"ur physikalische Chemie, Georg-August-Universit\"at G\"ottingen, Tammanstra\ss e 6, 37077 G\"ottingen, Germany}

\author{Pascal Larrégaray}
\affiliation{Universit\'e de  Bordeaux, 351 Cours de la Lib\'eration, 33405 Talence, France}
\affiliation{CNRS, 351 Cours de la Lib\'eration, 33405 Talence, France}

\author{Daniel J. Auerbach}
\affiliation{%
 Max-Planck-Institut f\"ur biophysikalische Chemie, Am Fa\ss berg 11, 37077 G\"ottingen, Germany}

\author{Dirk Schwarzer}%
\affiliation{%
 Max-Planck-Institut f\"ur biophysikalische Chemie, Am Fa\ss berg 11, 37077 G\"ottingen, Germany}

\author{Alec M. Wodtke}
\affiliation{%
 Max-Planck-Institut f\"ur biophysikalische Chemie, Am Fa\ss berg 11, 37077 G\"ottingen, Germany}
\affiliation{Institut f\"ur physikalische Chemie, Georg-August-Universit\"at G\"ottingen, Tammanstra\ss e 6, 37077 G\"ottingen, Germany}
\author{Alexander Kandratsenka}
\email{akandra@mpibpc.mpg.de}
\affiliation{%
 Max-Planck-Institut f\"ur biophysikalische Chemie, Am Fa\ss berg 11, 37077 G\"ottingen, Germany}



\date{\today}

\begin{abstract}
The Langevin equation 
includes a random force to maintain equilibrium and prevent friction from bringing motion to a standstill; but for ballistic motion, the random force is often neglected. Here, we use the Langevin equation for molecular dynamics simulations of 
$
2.76$\,eV H-atoms 
experiencing electronic friction in collisions with 300\,K metals, where a random force arises from thermal electron-hole pairs.   Simulations without the random force fail dramatically to reproduce experiment, although  
the incidence energy is much larger than $k_\text{B}T$
. We analyze the Ornstein-Uhlenbeck process to show that this is a general property of ballistic particles experiencing friction under the influence of thermal fluctuations. 

\end{abstract}

\maketitle



The Langevin equation originally served as an alternative to Einstein’s \cite{einstein08} and Smoluchowski’s \cite{von_Smoluchowski06} treatment of Brownian motion, the jittery back-and-forth hopping first seen under a microscope for pollen particles suspended in water that forms the physical basis for thermal diffusion. 
It explicitly describes  time-dependent fluctuations seen in experiments with a random force derived using the fluctuation-dissipation theorem \cite{langevin08}. The insights clarified by  the random force helped to establish the molecular view of matter \cite{coffey}. 
Today, the Langevin equation is the most common theoretical ansatz used to model electronically nonadiabatic effects between atoms (or molecules) and solid metals \cite{Schaich1974,dAgliano75,li_wahnstrom92a,head-gordon95,alducin17}. Here, nuclear motion takes the part of the Brownian pollen particle and thermally fluctuating electron-hole pairs ({\it ehp}) of the metal play the role of the jiggling water molecules. 

These frictional models of electronically nonadiabatic motion 
have broad applicability, for example, to describe the ion stopping power of metals \cite{fermi47,ritchie59,echenique1981,echenique1986,puska83}, nonadiabatic dynamics \cite{alducin17, alducin13, trail03, juaristi08, saalfrank14, novko15, blanco14, pena19, ibarguen20, jiang16} like the thermalization of hot atoms \cite{blanco14} and even the mechanism of hydrogen atom adsorption to metal surfaces \cite{buenermann15,Dorenkamp18}. 
Furthermore, a variety of models for electronic friction have been proposed \cite{puska83,hellsing84,head-gordon95,juaristi08,rittmeyer15,maurer16,askerka16,dou18,spiering19,zhang2020,gerrits20}, each based on different physical assumptions. Experimental tests of these models are needed to determine which are most reliable.

Inelastic H atom scattering from metal surfaces \cite{buenermann15,Dorenkamp18,jiang19} provides a direct probe of electronically nonadiabatic forces in a system that can be  treated classically in full dimensions, including surface atom motion \cite{janke15,janke13}. Experimental and theoretical energy-loss distributions can be compared to test models of electronic friction.  
But, since the Langevin equation describes how a system evolves under the influence of a frictional drag and a random force, the experimental manifestations of a model of electronic friction cannot be realized without the influence of the random force.  This poses the question:  how important is the influence of the random force? 

When the Langevin equation is used to describe diffusion, the random force is essential, preventing motion from eventually coming to a standstill due to friction. But to describe 
scattering and reactions of atoms and molecules at surfaces, its importance is not as obvious. In fact, the random force has often been ignored \cite{trail03,alducin17, alducin13, trail03, juaristi08, saalfrank14, novko15, blanco14, pena19, ibarguen20, jiang16,blanco14}, using as justification the fact that the projectile kinetic energy $\epsilon_0$ is much larger than thermal energy $k_\text{B} T$. On the face of it, this assumption appears reasonable. 
For example, should we wish to describe ballistic motion of an H atom in collisions with a metal, there is no danger of the system coming to a standstill and the timescale of a scattering collision can be very short, possibly rendering the {\it ehp} fluctuations unimportant. 

In this Letter, we present molecular dynamics simulations using the Langevin equation to describe H atom scattering from room temperature metal surfaces, where the incidence energy is large and where interactions last only $\approx25$\,fs. We compare these calculations to experimentally derived H atom energy loss distributions \cite{buenermann15,Dorenkamp18}. The trajectory simulations are generally in good agreement with experiment provided the random force is included. However, neglecting it produces energy loss distributions that qualitatively fail to describe the experimental ones, even for $\epsilon_0/k_\text{B} T > 100$. Only at surface temperatures below about 100\,K, does the influence of the random force diminish. This work shows that the physical mechanisms of nonadiabatic dissipation can easily be obscured by the random force. 

To investigate the influence of the random force in the Langevin equation we performed 
molecular dynamics (MD) simulations of H atom scattering from two metals, Au and W. We compared outcomes employing two different approaches: model I \cite{janke15}, where  atom-surface interaction is described by  a full-dimensional potential energy surface (PES) constructed by means of the Effective Medium Theory \cite{emt87,janke13, kammler17}, and the surface is represented by a slab of metal atoms with periodic boundary conditions; and model II \cite{Petuya14,Galparsoro2016}, where a 3D
PES produced by the Corrugation Reducing Procedure \cite{busnengo00, kresse00, olsen02}
is used, and the surface is described by a generalized Langevin oscillator \cite{adelman76,tully80,polanyi85}.
In both models the nonadiabatic coupling is described by the electronic friction coefficient depending on the background electron density (local density friction approximation) \cite{puska83, juaristi08}. In model I, the background electron density appears self-consistently as it is necessary to calculate the energy; it depends on the positions of both projectile and surface atoms \cite{janke15}. In model II, one has to do additional {\it ab initio} calculations with the frozen surface to get the electron density as a function of a projectile position \cite{Galparsoro2016}.     

The projectile is propagated by the Langevin equation of motion 
\begin{equation}
m{\bm {\ddot r}} = - \frac{\partial E}{\partial {\bm r}} - m\eta_{\text{el}}
{\bm {\dot r}} + \bm F_{\text L}(t),
\end{equation}
where $E$ is the potential energy of the system, $m$ is the mass of the projectile,  $\eta_{\text{el}}$ is the electronic friction coefficient dependent on the system's geometry, and $\bm F_{\text L}(t)$ is the random force, which follows a Gaussian distribution with zero average
and variance determined by the fluctuation-dissipation theorem \cite{kubo}
\begin{equation}\label{eq:fdt}
\left\langle {\bf F}_{\text{L}} (t){\bf F}_{\text{L}} (t^\prime)\right\rangle =  2k_{\text{B}}T_{\text{el}} m \eta_{\text{el}}{\bf I}
\delta (t-t^\prime),
\end{equation}
where $T_{\text{el}}$ is the temperature of the electron bath, and ${\bf I}$ is the 3D unity matrix. 


 
For both models, trajectories were run with an incidence energy $\epsilon_0=2.76$ eV and an incidence angle $\vartheta_{\text{i}}=45^{\circ}$ with respect to the surface normal. The azimuthal angles $\varphi_{\text{i}}$ for the gold and tungsten calculations were defined with respect to the $\left[ 10\bar{1}\right]$ direction and $\left[ 001\right]$ direction, respectively. Trajectories were initiated with the projectiles placed at random lateral positions 6\,$\text{\AA}$ above the surface. The calculations were stopped after 1\,ps or if the scattered projectile was found more than 6.05\,$\text{\AA}$ above the surface. 

\begin{figure}[hbt!]
\includegraphics[width=0.5\textwidth]{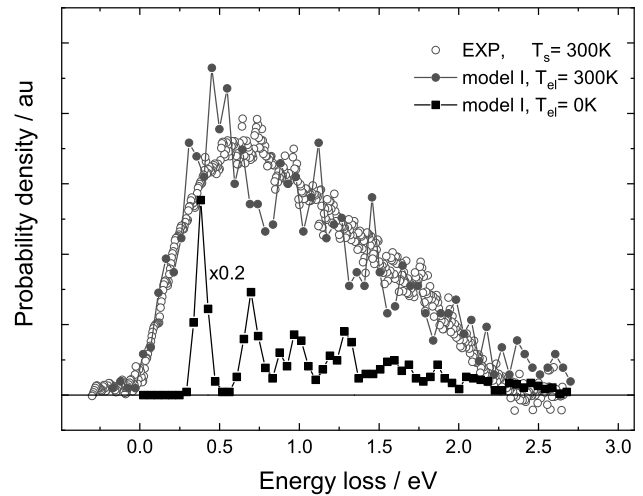}
\caption{\label{fig:comp_exp} 
H atom inelastic scattering from Au(111): Comparing theory and experiment.
Using model I with $T_\text{el}=300$\,K  ($\bullet$) good agreement with experiment ($\circ$) is found. By setting $T_\text{el}=0$\,K, the random force is deactivated and theory ($\filledsquare$) deviates from experiment. For all three curves, $\epsilon_0=$ 2.76 eV, phonon temperature is $300$\,K, $\vartheta_{\text{i}}  = 45^{\circ}$ and $\vartheta_{\text{s}}  = 45^{\circ}$ with respect to the surface normal while  $\varphi_{\text{i}}=0^{\circ}$ with respect to the $\left[ 10\bar{1}\right]$ direction. Experimental results are taken from 
\cite{buenermann15}.}
\end{figure}

Figure \ref{fig:comp_exp} shows results using model I. The energy loss distribution constructed from the MD trajectories ($\bullet$) that scatter into angles that match the angular acceptance of the experiment  successfully reproduces an experimentally obtained energy loss distribution $(\circ)$. 
The scattered H atoms exhibit a mean energy loss of approximately 1\,eV and appear in a distribution with a remarkably broad width of 2.5\,eV due to energy exchange with {\it ehp} and phonons. 
When $T_\text{el}$ is set to $0$\,K, the MD simulations $(\filledsquare)$ fail spectacularly. Note that setting $T_{\text{el}}=0$\,K is equivalent to neglecting the random force.

\begin{figure}[hbt!]
\centering
\includegraphics[width=0.48\textwidth]{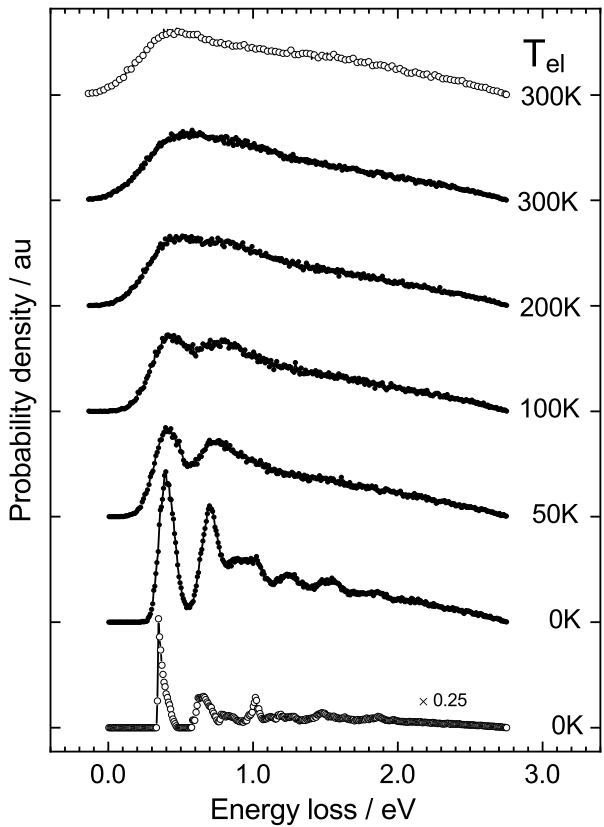}
\caption{\label{fig:comps} 
Electronic temperature determines the shape of the energy loss distribution.
Energy loss distributions are shown for scattered H atoms from a moving Au(111) surface  with a phonon temperature of $300$\,K ($\bullet$) and with a static lattice approximation ($\circ$) at various electronic temperatures  $T_\text{el}$. Incidence conditions are the same as in Fig.~\ref{fig:comp_exp}; but here, trajectories at all scattering angles are used. } 
\end{figure}

We show the influence of the random force on the energy loss distribution in Figure~\ref{fig:comps}.
Here, MD trajectories are generated as in Fig.~\ref{fig:comp_exp} using a PES with moving Au atoms ($\bullet$), but $T_{\text{el}}$ is varied between 300 and 0 K.  As $T_{\text{el}}$ decreases, peaks appear in the energy loss distribution. Analysis of trajectories reveals that these peaks correspond to 
"bounces", {\it i.\,e.}, to 
interactions involving different number of collisions between H and Au atoms. The energy loss increases approximately linearly with each additional collision, reflecting the increased  interaction time.   
Also shown in Fig.~\ref{fig:comps} are two MD calculations employing a frozen surface 
($\circ$) with $T_{\text{el}}=0$\,K and  300\,K. For $T_{\text{el}}=0$\,K, peaks are even sharper than for the moving surface MD simulations at $T_{\text{el}}=0$\,K, the difference reflecting kinetic energy exchange between H and Au atoms. In contrast, at $T_{\text{el}}=300$\,K it is hard to distinguish the energy loss distribution obtained when using a static surface approximation from that obtained when Au atoms are allowed to move. 

\begin{figure}[hbt!]
\centering
\includegraphics[width=0.45\textwidth]{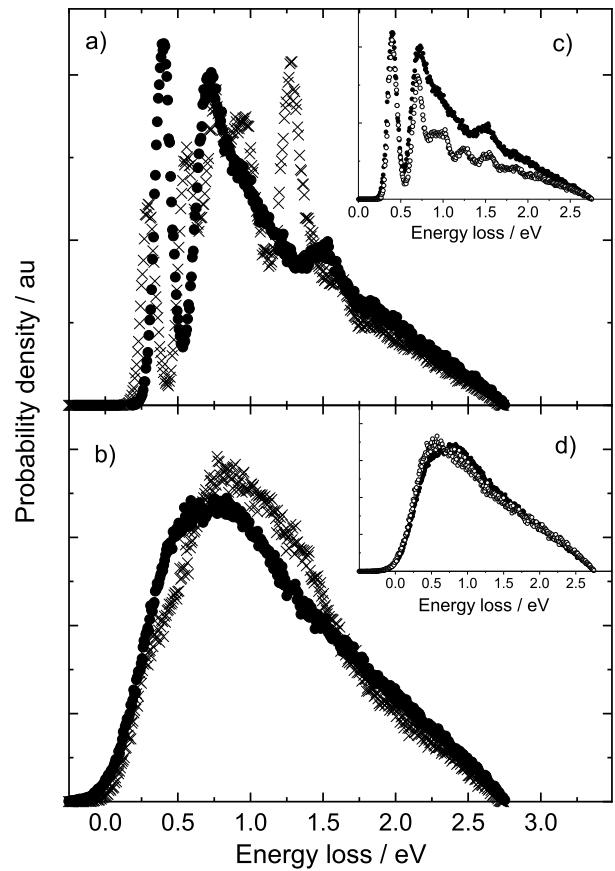}
\caption{\label{fig:EMT_CRP} 
The obscuring influence of the random force at modest temperature:
angle-integrated energy loss distributions for scattered H atoms from W(110) using model I ($\bullet$) and II ($\times$) at (a) $T_\text{el}= 0$\,K and (b) $T_\text{el}= 300$\,K;   in the insets 
the energy loss spectra for H scattering from W(110) ($\bullet$) and Au(111) ($\circ$) are compared at (c) $T_\text{el}= 0$\,K and (d) $T_\text{el}= 300$\,K using model I.  The phonon temperature in all cases is $300$\,K. 
}
\end{figure}

We next investigate the sensitivity of the energy loss distribution 
to the choice of the dynamical model. Figs.~\ref{fig:EMT_CRP}a and \ref{fig:EMT_CRP}b show
the results of MD simulations for
H scattering from W(110) computed with models I and II. In Fig.~\ref{fig:EMT_CRP}a, where $T_\text{el}=0$\,K, the two energy loss distributions are clearly distinguishable from one another. This is, however, not the case when $T_\text{el}=300$\,K (Fig.~\ref{fig:EMT_CRP}b). Despite the moderate temperature and high H atom incidence energy, it is clear that the broadening effects of the random force on the energy loss distribution smear out the differences in the scattering dynamics resulting from the two models. 

It is noteworthy that  similar effects were observed  in studying adiabatic dynamics of Ar and Xe at metal surfaces \cite{tully80}, where friction and fluctuating (random) force were due to the thermal bath of phonons. Here, the calculated properties (sticking coefficients {\it etc.}) were not sensitive to the details of atom-surface interaction or changes in the phonon spectral density. 

The sensitivity of the energy loss distribution to the identity of the metal is also interesting. To study this we compared MD scattering results from two metals. Figs.~\ref{fig:EMT_CRP}c and \ref{fig:EMT_CRP}d show comparisons of H scattering from {\it fcc} Au(111) ($\circ$) and {\it bcc} W(110)($\bullet$), both using model I. Remarkably, the energy distributions associated with these two metals can only be distinguished at low electronic temperature.   

To better understand  the surprisingly strong influence of the random force on the width of energy loss distributions, consider a closely related problem that has an analytical solution: the one-dimensional motion of an ensemble of  particles of mass $m$ with incidence energy $\epsilon_0$ subjected to friction with characteristic deceleration time $\tau$ experiencing a random force at temperature $T$. The random force distribution is Gaussian with the second moment 
defined by Eq.~\eqref{eq:fdt} and the friction coefficient $\eta_\text{el}=\tau^{-1}$ is constant. This is known as the Ornstein-Uhlenbeck (OU) process \cite{ornuhl} and we can use it to describe a scattering trajectory that has not reached equilibrium.

The ensemble's initial velocity distribution is $\delta(v-v_0)$;  thereafter it is normal, 
with  the time-dependent  expectation $\bar v(t)$ and standard deviation $\sigma_v(t)$
given by 

\begin{equation}
  \bar v(t)= v_0e^{-t/\tau},\qquad
  \sigma_v(t)=\sqrt{\frac{k_\text{B}T}{m}\xi(t)},
\end{equation}
where $\xi(t)= 1-e^{-2t/\tau}$ and $v_0=\sqrt{2\epsilon_0/m}$ \cite{vanKampen,Risken84}.
From this, we derived  
the time-dependent energy distribution
\begin{equation}
\label{eq:edistr}
P(\epsilon,t)= \frac{e^{-(\epsilon+\epsilon_0 e^{-2t/\tau})/k_\text{B}T\xi(t)}}{\sqrt{\pi\epsilon k_\text{B}T\xi(t)}}
  \cosh\frac{2e^{-t/\tau}\sqrt{\epsilon_0\epsilon}}{k_\text{B}T\xi(t)},
\end{equation}
which has the form of a folded normal distribution \cite{leone61}.

\begin{figure}[hbt!]
\centering
\includegraphics[width=0.45\textwidth]{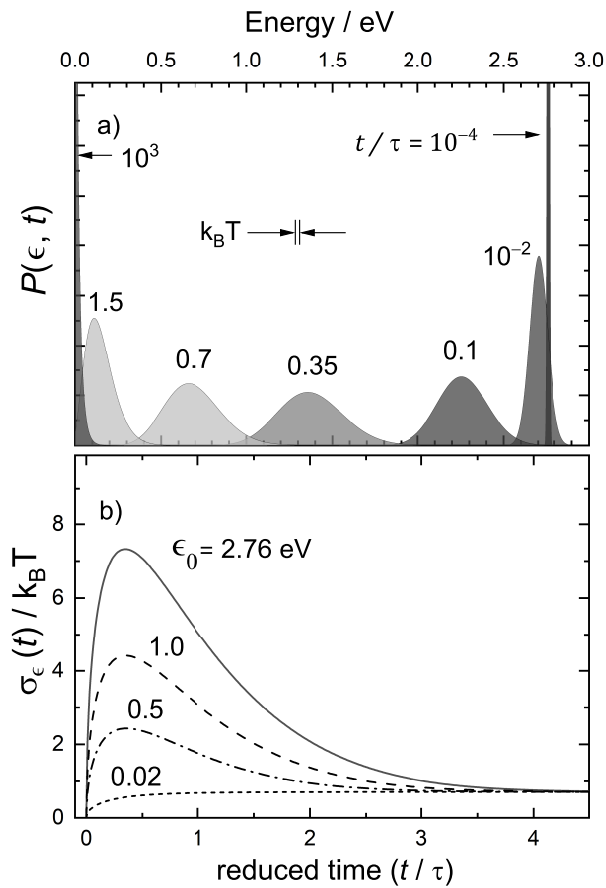}
\caption{
\label{fig:ou}
Time-dependent energy distribution of the Ornstein-Uhlenbeck process. 
(a) A particle with  incidence energy $\epsilon_0=$ 2.76 eV decelerates under a frictional drag subject to thermal fluctuations at $T=$ 300 K. Energy distributions are shown at various times, in units of $\tau$, the characteristic time for deceleration. (b) The width of the distribution is shown for various choices of incidence energy $\epsilon_0.$}
\end{figure}

Figure \ref{fig:ou}a shows energy distributions from Eq.~\eqref{eq:edistr}  for the one-dimensional OU process at $\epsilon_0=2.76$\,eV and $T=300$\,K. Figure \ref{fig:ou}b shows the time-dependent width of  the energy distribution 
\begin{equation}
\label{eq:sigt}
\sigma_\epsilon(t)=\frac{k_\text{B}T\xi(t)}{\sqrt{2}}\sqrt{1+\frac{4\epsilon_0}{k_\text{B}T}\frac{1-\xi(t)}{\xi(t)}}
\end{equation}
for four choices of $\epsilon_0$ and at $T=300$\,K. At $t=0$ the energy distribution is a delta-function. 
At intermediate time, $\sigma_\epsilon(t)$ overshoots $k_\text{B} T$ reaching a maximum given by
\begin{align}\label{eq:sigmax}
 \sigma_\epsilon(t_\text{max}) = \sqrt{\frac{2 k_\text{B}T \epsilon_0^2}{4\epsilon_0 - k_\text{B}T}},
\end{align}
where 
\begin{align}\label{eq:tmax}
t_\text{max}=\frac{\tau}{2} \ln\frac{4\epsilon_0-k_\text{B}T}{2\epsilon_0-k_\text{B}T},
\end{align}
before eventually falling back to the equilibrium value $k_\text{B} T/\sqrt{2}$ in the limit of infinite time.  Under the conditions of Figure \ref{fig:ou}a, $t_\text{max}=0.35\tau$, but $\sigma_\epsilon(t)$ is much larger than $k_\text{B}T$ already at $t= 0.1 \tau$ and remains quite broad until nearly completely decelerated.

A naive view of Eq.~\eqref{eq:fdt} might suggest that because the distribution of random forces scales as $\sqrt{k_\text{B} T}$, the width of the energy distribution scales similarly. However, when the random force introduces a thermally distributed change in velocity $\delta v$, the resulting change in energy scales as $(v_0+\delta v)^2-v_0^2 = 2v_0\delta v+\delta v^2$. The term $2v_0\delta v$ contributes to the energy distribution width in proportion to the hyperthermal velocity $v_0$.
Eq.~\eqref{eq:sigmax} shows this; $\sigma_\epsilon(t_\text{max})$ scales as
 $\sqrt{\epsilon_0 k_\text{B} T}$ for $\epsilon_0\gg k_\text{B}T$.
Furthermore, Eq.~\eqref{eq:tmax}  shows that the thermal overshoot in the width of the energy distribution is absent only when $\epsilon_0<k_\text{B}T/2$ (see also Fig. \ref{fig:ou}b). Clearly, for the OU process one cannot justify ignoring the influence of the random force with an argument that $\epsilon_0$ is much larger than $k_\text{B}T$. 
It is not then surprising that this argument is also incorrect when computing nonadiabatic MD trajectories in many dimensions.  

The observations arising from our analysis of the H atom energy loss distributions and of the OU process suggest that neglecting the random force for ballistic motion is generally unwise. 
The results of this work also serve a warning. The generally good agreement seen between H atom scattering experiments and MD simulations with electronic friction is due largely to broadening effects introduced by the random force. To experimentally distinguish different theories of nonadiabatic dynamics, experiments at low surface temperature are needed. This could put new demands on theory, as quantum dynamics may be important at low temperature. 

\begin{acknowledgments}
We would like to thank Prof.~John C.\ Tully for helpful discussions and comments.
This work was partly funded by the Deutsche Forschungsgemeinschaft (DFG, German Research Foundation) - 217133147/SFB 1073, project A04. AK acknowledges European Research Council (ERC) under the European Union's Horizon 2020 research and innovation programme (grant agreement no. 833404). RMB and PL acknowledge funding from the French Embassy in Cuba and the Transnational Common Laboratory QuantumChemPhys: Theoretical Chemistry and Physics at the Quantum Scale (ANR-10-IDEX-03-02)
\end{acknowledgments}


\providecommand{\noopsort}[1]{}\providecommand{\singleletter}[1]{#1}%

\end{document}